\documentclass{article}

\usepackage[preprint]{neurips_2026}

\usepackage[utf8]{inputenc} 
\usepackage[T1]{fontenc}    
\usepackage[colorlinks]{hyperref}
\usepackage{url}            
\usepackage{booktabs}       
\usepackage{amsfonts}       
\usepackage{nicefrac}       
\usepackage{microtype}      
\usepackage{xcolor}         
\usepackage{placeins}         

\usepackage{bm}
\usepackage{siunitx}
\usepackage{stfloats}
\usepackage{tikz}
\usepackage{comment}
\usepackage{amsmath}
\usepackage{amssymb}
\usepackage{mathtools}
\usepackage{amsthm}
\usepackage{dsfont}
\usepackage{wrapfig}
\usepackage{microtype}
\usepackage{multirow}
\usepackage{graphicx}
\usepackage[capitalize,noabbrev]{cleveref}
\usepackage{algorithm}
\usepackage{algorithmic}
\usepackage{enumerate}
\usepackage{xcolor,colortbl}
\usepackage{nicefrac}
\usepackage[most]{tcolorbox}
\usepackage{enumitem}

\definecolor{mydarkblue}{rgb}{0,0.08,0.45}
\hypersetup{
  colorlinks=true,
  linkcolor=mydarkblue,
  citecolor=mydarkblue,
  urlcolor=mydarkblue
}

\definecolor{darkgreen}{rgb}{0.0, 0.5, 0.0}

\newcommand{\Dcal}{\mathcal{D}}
\newcommand{\wh}{\widehat}
\DeclareMathOperator{\PPD}{PPD}

\usepackage{amsmath,amsfonts,bm}

\def\1{\bm{1}}

\def\eps{{\epsilon}}

\DeclareMathAlphabet{\mathsfit}{\encodingdefault}{\sfdefault}{m}{sl}
\SetMathAlphabet{\mathsfit}{bold}{\encodingdefault}{\sfdefault}{bx}{n}

\newcommand{\E}{\mathbb{E}}

\newcommand{\R}{\mathbb{R}}

\theoremstyle{plain}

\newtheorem{proposition}{Proposition}[section]

\newtheorem{corollary}{Corollary}[section]
\theoremstyle{definition}

\theoremstyle{remark}

\usepackage[textsize=tiny]{todonotes}

\title{Uncertainty Quantification for Prior-Data Fitted Networks using Martingale Posteriors}

\author{%
  Thomas Nagler \qquad David R\"ugamer
    \\[10pt]
  Department of Statistics, LMU Munich\\
  Munich Center for Machine Learning (MCML) \\
  \texttt{\{t.nagler,david.ruegamer\}@lmu.de} \\
}

\begin{document}

\maketitle

\begin{abstract}
  Prior-data fitted networks (PFNs) have emerged as promising foundation models for prediction from tabular datasets, achieving state-of-the-art performance on small to moderate data sizes without tuning. While PFNs are motivated by Bayesian ideas, they do not provide any uncertainty quantification for predictive means, quantiles, or similar quantities. We propose a principled, efficient, and tuning-free sampling procedure to construct Bayesian posteriors for such estimates based on martingale posteriors, and prove its convergence.
    Several simulated and real-world data examples showcase the efficiency and calibration of our method in inference applications.
\end{abstract}

\section{Introduction}

Prior-data fitted networks (PFNs) are foundation models \citep{hollmann2023tabpfn,mueller2022transformers} that allow for in-context learning, i.e., the ability to learn at inference time without any parameter updates \citep{garg2022can}. TabPFN, a transformer pre-trained on synthetic data for in-context learning on tabular datasets, has recently attracted a lot of interest. 
TabPFN \citep{hollmann2023tabpfn, hollmann2025accurate} and related variants such as TuneTables \citep{feuer2024tunetables}, LocalPFN \citep{thomas2024retrieval}, or TabICL \citep{qu2025tabicl, qu2026tabiclv2} have been shown to achieve state-of-the-art performance on tabular benchmarks by pre-training on purely synthetic data. Since PFNs and extensions learn in-context, there is no need for further model (fine-)tuning on the inference task. 


Recent extensions of PFNs allow their applicability to large datasets \citep{feuer2024tunetables}, the use of PFN ``priors'' for latent variable models \citep{reuter2025transformerslearnbayesianinference}, and simultaneously minimizing bias and variance to improve their performance \citep{liu2025tabpfn}. PFNs are also related to simulation-based inference and amortized inference, but have slightly different goals and do not amortize across a single but multiple datasets \citep{reuter2025transformerslearnbayesianinference}.
While introduced as a Bayesian method and approximation to the posterior predictive, PFNs can also be interpreted as pre-tuned untrained predictors \citep{nagler2023statistical}. This also relates to the question of what uncertainty PFN models can provide.



PFNs approximate the posterior predictive distribution for the label given some feature values. Despite the name, this only yields point estimates of the most relevant predictive quantities, such as the conditional mean or quantiles.
Due to the complex nature of PFNs, it is difficult to assess the uncertainty of these point estimates. This explains why, despite the practical relevance, methods for such uncertainty assessment are currently lacking.



In this article, we propose a principled and efficient method to construct Bayesian posteriors for such estimates using the idea of \emph{Martingale Posteriors} \citep[MPs;][]{fong2023martingale}.

\newpage

\paragraph{Our contributions}
\begin{enumerate}[leftmargin=1.5em]
    \item We introduce a formulation of the MP framework for inference of predictive quantities conditional on a specific feature value $x$.
    \item We propose an efficient, nonparametric resampling scheme yielding an approximate posterior for the point estimates derived from a PFN, and prove its convergence.
    \item We adapt existing learning rate schedules and discuss the role of contraction rates in light of nonparametric PFN estimators.
    \item We illustrate the new method in several simulated and real-world data applications.
    \item We perform a variety of ablation studies providing insights into the driving factors of our proposal's efficacy and failure modes of alternative methods.
\end{enumerate}

Our work provides an essential tool for principled inference with the increasingly popular PFN methods. Our approximate martingale posterior (AMP) algorithm provides an uncertainty quantification layer that aligns perfectly with the strengths of PFNs: a tuning-free method yielding well-calibrated credible intervals in a matter of seconds.

\paragraph{Related work} There are two works that are closely related to ours, but target different quantities. \citet{ng2025tabmgpmartingaleposteriortabpfn} propose martingale posteriors for unconditional, rather than predictive summaries of the data. \citet{fortini2026principled} propose a Gaussian approximation of the joint posterior of a finite number of event probabilities, but do not provide uncertainty estimates for, e.g., conditional means or quantiles considered in this article.
Additionally, the corresponding algorithms require many repeated calls to the PFN, making them orders of magnitude slower than ours.

\begin{figure*}[t]
\centering
\includegraphics[width=0.97\linewidth]{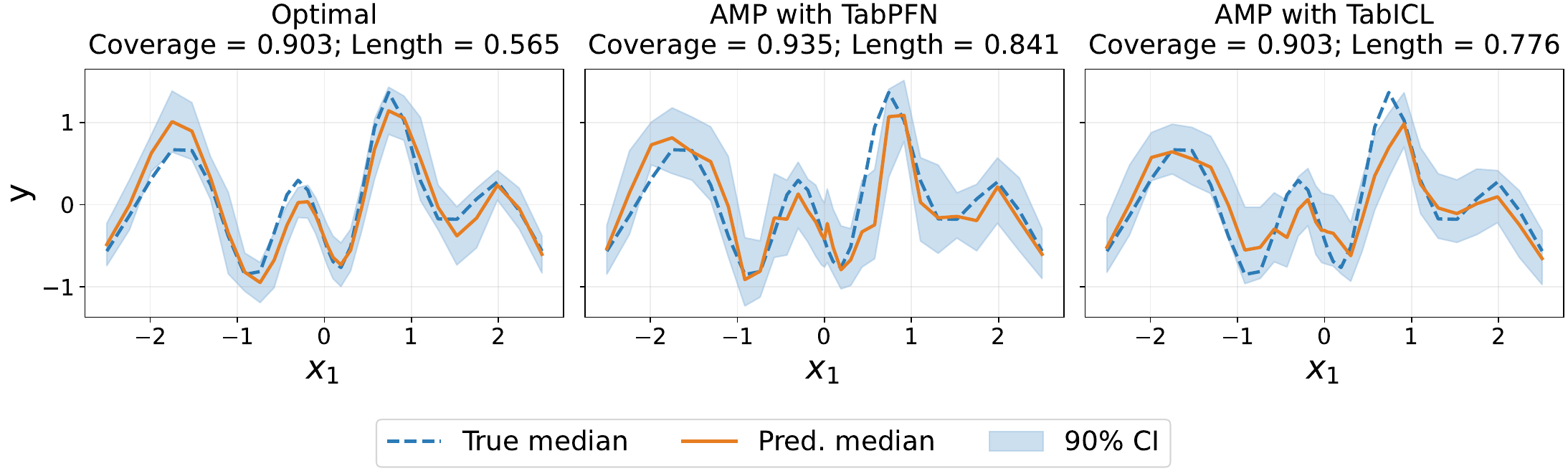}
    \caption{Comparison of estimated median function (orange line) and credible intervals (shaded blue area) for a non-linear feature effect (dashed blue line) in a Bayesian additive model based on different methods (facets). Numbers below each method state the average coverage and interval length across all data points. ``Optimal'' corresponds to the posterior induced by the chosen prior and likelihood, which in this case can be analytically derived (see \cref{sec:sim} for details). Our approach (AMP) can be defined for any PFN that allows access to the learned predictive distribution (here, TabPFN and TabICL).}
    \label{fig:1}
\end{figure*}

\section{Background}

We consider a tabular prediction task with labels $y \in \R$ and features $x \in \R^d$ drawn from a joint distribution $P$. A typical problem in such tasks is to estimate predictive quantities such as conditional means $\E[y | x]$, 
conditional probabilities $P(y | x)$, or conditional quantiles $P^{-1}(\alpha | x)$. Because the true distribution $P$ is unknown and only a finite amount of data $\Dcal_n = (y_i, x_i)_{i = 1}^n$ is available, estimates of such quantities bear some uncertainty. Our goal is to quantify this uncertainty.

\subsection{Prior-data fitted networks}

Prior-data fitted networks are foundation models trained to approximate the posterior predictive density 
$\PPD(y| x) = p(y | x, \Dcal_n)$,
which quantifies the likelihood of observing label $y$ given that the feature is $x$ and $\Dcal_n$ has been observed. The PPD is a Bayesian concept and implicitly involves a prior over the distributions $P$ that could have generated the data. To approximate the PPD with a PFN, 
a deep neural network---typically a transformer---is pre-trained on simulated datasets with diverse characteristics.  After pre-training, the network weights are fixed, and the approximate PPD for a new training set can be computed through a single forward pass without additional training or tuning.  

The PPD quantifies uncertainty about the label $y$. However, it mixes the aleatoric and epistemic components of uncertainty \citep{hullermeier2021aleatoric}. From the PPD alone, it is impossible to disentangle these parts. Consequently, PFNs do \emph{not} provide uncertainty estimates for predictive summaries, such as conditional mean, probabilities, and quantiles.

\subsection{Bayesian inference}

In classical Bayesian inference, the set of possible distributions $P = P_\theta$ is indexed by some parameter $\theta$. A \emph{prior} distribution $\pi(\theta)$ is elicited to quantify our beliefs about the likelihood of the possible values of $\theta$ before seeing any data. 
After observing $\Dcal_n$, this belief is updated to a \emph{posterior} $\pi(\theta | \Dcal_n)$ of the parameter $\theta$ given the data. For predictive inference, the PPD can be computed as 
\begin{align*}
    \PPD(y | x) = \int p_\theta(y | x)  \pi(\theta | \Dcal_n) \, d\theta.
\end{align*}
In contrast to the PPD, the posterior $\pi(\theta | \Dcal_n)$ also quantifies uncertainty for other interest quantities. For example, the posterior distribution for the conditional mean $\mu(x) = \int y\,  p_\theta(y | x) dy$ is given by 
\begin{align*}
    \Pi(\mu(x) \in A) = \int  \1\left\{\int y\, p_\theta(y | x) dy \in A\right\} 
   \pi(\theta | \Dcal_n)\, \,d\theta,
\end{align*}
for any $A \subseteq \R$.
PFNs neither provide an explicit model for $p_\theta$ nor an explicit prior $\pi(\theta)$, although both may be implicit in the PPD. The following shows how Bayesian posterior inference can be approached when only the PPD is available.

\subsection{Martingale posteriors}

Martingale posteriors were recently introduced by \citet{fong2023martingale} as a new method for Bayesian uncertainty quantification. Its core idea is to reverse the direction of the Bayesian inference. In classical Bayesian inference, the posterior is derived from a prior and likelihood, which then implicitly leads to the PPD. MP inference starts from the PPD and leaves the prior $\pi(\theta)$ implicit. An appropriate sampling scheme and Doob's theorem then allow us to derive posteriors for virtually all quantities of interest (e.g., the conditional mean $\mu(x)$). 

To simplify our outline of the approach, consider the case where there are no features, and we are interested in unconditional inference. An extension to our predictive inference setting will be made explicit in \cref{sec:mp_conditional}. Suppose we have observed data $y_{1:n} = (y_1, \dots, y_n)$.

The MP approach involves iteratively sampling  
\begin{align} \label{eq:iterproc}
    y_{n + 1} \sim p(\cdot | y_{1:n}), \quad  y_{n + 2} \sim p(\cdot | y_{1:(n+1)}), \quad y_{n + 3} \sim p(\cdot | y_{1:(n+2)}), \quad \dots
\end{align}
$N$ times, which yields a sample $y_{(n + 1):(n + N)}$ drawn from the predictive joint distribution 
\begin{align*}
    p(y_{(n + 1):(n + N)}| y_{1:n}) = \prod_{i = 1}^N p(y_{n + i}| y_{1:(n + i - 1)}).
\end{align*}
Observe, however, that the samples are not independent. As a consequence, the long-run empirical distribution of the obtained sample,
\begin{align*}
    F_\infty(y) = \lim_{N \to \infty} \frac 1 N \sum_{i = 1}^N \1(y_{n + i} \le y),
\end{align*}
is a random function and comes out differently whenever the sampling procedure is repeated. 
Denote by $\Pi(F_\infty | \Dcal_n)$ the distribution of this function (which depends on the data $\Dcal_n$ we start with).
For any parameter $\theta = \theta(P)$ of interest, the martingale posterior is now given as 
\begin{align*}
     \Pi(\theta \in A | \Dcal_n) = \int \1\{\theta(F_\infty) \in A\} d\Pi(F_\infty | \Dcal_n),
\end{align*}
where $A$ is any Borel set on the space where the parameter $\theta$ lives.
Furthermore, Doob's theorem \citep{doob1949application}  implies that $\Pi(\theta  | \Dcal_n) $ coincides with the classical Bayes posterior for the prior $\pi(\theta)$ implicit in the PPD \citep{fong2023martingale}.

\section{Efficient martingale posteriors for prior-data fitted networks} \label{sec:main}

Martingale posteriors allow for Bayesian inference directly from the PPD. PFNs approximate the PPD, so using PFNs to construct a martingale posterior seems natural. However, there are two problems. 
First, \citet{falck24} found that modern transformer-based models substantially deviate from the \emph{martingale property}
\begin{align*}
    \E[p(y|y_{1:(n + k)}) | y_{1:n}] = p(y|y_{1:n}).
\end{align*}
Second, modern PFNs are based on transformer architectures that require $\Omega(n^2)$ operations per forward pass on a training set of size $n$. Iteratively computing $p(y| y_{1:(n + k)})$ for $k = 1, \dots, N$ thus has complexity $\Omega(N^3)$, which is prohibitive.



We propose to use the PPD implied by the PFN only as a starting point for the MP sampling scheme. After obtaining the PFN's PPD, we then iteratively update it using a nonparametric forward sampling scheme that 
ensures the martingale property. The resulting martingale posterior effectively treats the PFN output as a strong, informed prior, without requiring the PFN to provide coherent posterior updates or incurring computational overhead from iterative model evaluations.

\subsection{Martingale posteriors for conditional inference} 
\label{sec:mp_conditional}

We extend the unconditional sampling scheme outlined in the previous section to the conditional inference setting. \citet{fong2023martingale} already proposed one such extension. Their scheme involves forward sampling of the features $x_{(n+1):(n + N)}$. The distribution of the features isn't of primary interest, but it complicates the sampling procedure and slows its convergence exponentially, making it impractical beyond about 5 feature dimensions. To alleviate this, we propose to sample only the labels $y_{(n +1):(n + N)}$ conditional on the event that $x_{n + k} = x$, for a fixed value of $x$ and all $k = 1, \dots, N$.

Set $x_{n + k} = x$ for all $k \geq 1$, and define
\begin{align*}
    p_{k}(y) = p(y_{n + k + 1} | y_{1:(n + k)}, x_{1:(n + k)}),
\end{align*}
and $P_k$ as the corresponding CDF.
Applying Bayes' rule recursively gives
\begin{align*}
    p(y_{(n+1):(n+N)} | x_{(n + 1):(n +N)} = x, \Dcal_n) = \prod_{k = 0}^{N - 1} p_{k}(y_{n + k + 1}),
\end{align*}
which suggests that we can iteratively sample
\begin{align*}
    y_{n + 1} \sim P_0, \quad y_{n + 2} \sim P_1, \quad y_{n + 3} \sim P_2, \quad \dots.
\end{align*}

Denote the long-run empirical distribution of the obtained sample by
    $F_{\infty, x}(y) = \lim_{N \to \infty} N^{-1} \sum_{i = 1}^N \1(y_{n + i} \le y)$,
which is again a random function, even in the limit. Repeating the iterative sampling procedure gives us its distribution $\Pi(F_{\infty, x} | \Dcal_n)$.
For any conditional parameter $\theta(x) = \theta(P( \cdot \, | x))$ of interest, the martingale posterior is now given as 
\begin{align*}
     \Pi(\theta(x) \in A | \Dcal_n) = \int \1\{\theta(F_{\infty, x}) \in A\} d\Pi(F_{\infty, x} | \Dcal_n).
\end{align*}
Common examples of the parameter $\theta(x)$ are the conditional mean 
    $\theta(x) =\int y \, dP(y | x) dy$
or a conditional $\varphi$-quantile 
    $\theta(x) = P^{-1}(\varphi | x)$.

\subsection{Efficient PPD updates based on the Gaussian copula}

Observe that $p_0(y) = p(y | x, \Dcal_n)$ is the PPD approximated by the PFN.
However, the following update distributions $p_1, p_2, \dots$ are generally intractable. To alleviate this, we propose the following computationally efficient surrogate updates:
\begin{equation}
\begin{aligned} \label{eq:gc_update}
    P_{k}(y) = &(1 - \alpha_{n + k -1})P_{k - 1}(y) 
    + \alpha_{n + k -1} H_\rho(P_{k - 1}(y), P_{k - 1}(y_{n + k})),
\end{aligned}
\end{equation}
where $P_k$ is the CDF corresponding to $p_k$, $\alpha_{i}$ a learning rate, and
$H_\rho(u, v) = \Phi((\Phi^{-1}(u) - \rho \Phi^{-1}(v))/{\sqrt{1 - \rho^2}}),$
with $\Phi$ the standard normal cumulative distribution function.

The updates have a similar form to those proposed in the unconditional context by \citet{fong2023martingale}, who derived them a nonparametric density estimator
based on Dirichlet Process Mixture Models (DPMMs) and a copula decomposition of the conditional $p_k$. We propose several modifications to their procedure in the following sections to make it amenable to PFN inference.

The procedure has a hyperparameter $\rho$, corresponding to a bandwidth that smoothes the updates. 
For $\rho \to 1$, the $H_\rho$-term converges to the CDF of a Dirac measure at $y_{n + k}$. The update then amounts to sampling from $P_{k - 1}$ and then mixing $P_{k - 1}$ with this measure proportional to $a_{n + k}$. This shows similarities to the Bayesian bootstrap, although starting from a different $P_0$.
When $\rho < 1$, the Dirac measure is smoothed on the scale of probability integral transforms.
\citet{fong2023martingale} proposed to tune $\rho$ by maximizing the likelihood of the updated densities over the observed data. This is mainly to appropriately tune their initial estimate of the PPD. Since we use a different initial PPD, it is more appropriate to remain agnostic and choose $\rho \approx 1$, and our experiments use $\rho = 0.99$. 


\citet{fong2023martingale} suggested a learning rate $\alpha_i \sim (i+1)^{-1}$ for the unconditional context, where distribution functions can be estimated with convergence rate $n^{-1/2}$. As we shall see in \cref{prop:P_N_convergence}, the learning rate $\alpha_i$ determines the contraction rate of the posterior.  Our setting, however, is conditional on $x \in \R^d$, where nonparametric methods such as PFNs must converge more slowly, with rates depending on the dimension of the estimation problem \citep[see, e.g.,][]{stone1982optimal}. 
To accommodate such settings, we use a more general parameterization of the learning rate: 
\begin{equation} \label{eq:lr_para}
    \alpha_{i} = 2^\beta(i + 1)^{-\beta}, \quad  \beta > 1/2.
\end{equation}
The factor $2^\beta$ is chosen such that $\alpha_i = 1$ if only a single observation has been observed $(n = 1)$. The choice of $\beta$ is discussed in Section \ref{sec:rolebeta}.


\subsection{Theoretical properties} \label{sec:theory}

Despite the simplicity of the updates given in (\ref{eq:gc_update}), they provide essential theoretical guarantees. 
The following is a direct consequence of Theorem 3 by \citet{fong2023martingale}.
\begin{proposition} \label{prop:exchangeability}
It holds $(y_{N + 1}, y_{N + 2}, \dots) \to_d (z_1, z_2, \dots)$ as $N \to \infty$ where $(z_1, z_2, \dots)$ has an exchangeable distribution. 
\end{proposition}
By de Finetti's theorem (e.g., Theorem 1.49 in \citealp{schervish2012theory}), it then follows that there is a random variable $\Theta$ such that $(z_1, z_2, \dots)$ is conditionally \emph{iid} given $\Theta$. The distribution of $\Theta$ can be interpreted as an implicit prior. In our setting, this prior depends both on the initial PPD implied by the PFN and the Gaussian copula updates specified by (\ref{eq:gc_update}). In particular, the following result follows from \cref{prop:exchangeability} above and Theorem 2.2 of \citet{berti2004limit}.
\begin{proposition} \label{prop:existence}
 Suppose that $P_0$ is absolutely continuous with respect to the Lebesgue measure. 
 Then there exists a random probability distribution $P_{\infty, x}$ such that $\lim_{N \to \infty} P_{N}(y) = P_{\infty, x}(y)=  F_{\infty, x}(y)$ almost surely.
\end{proposition}
The proposition implies that the (random) limit $P_{\infty, x}$ is well defined and that the iterative sampling scheme is a valid way to draw from its distribution.
We can be more precise:
\begin{proposition}  \label{prop:P_N_convergence}
     For any $y \in \R$, $\beta > 1/2$,  the following holds  with probability at least $1 - \delta$:
     \begin{align*}
        |P_{\infty, x}(y) - P_{0}(y)| \le \sqrt{\frac{2^{1 + 2\beta} \log(2/\delta)}{2\beta - 1}}n^{-\beta + 1/2}.
    \end{align*}
\end{proposition}
The proof, given in \cref{sec:proof}, is based on a time-uniform version of the Azuma-Hoeffding concentration inequality \citep{howard2020time}, the martingale property of the copula updates, and the learning rate schedule. 
Importantly, time-uniform martingale concentration allows us to directly bound the deviation of interest $|P_{\infty, x} - P_0|$ as opposed to $|P_M - P_0|$ for fixed $M < \infty$ as in \citet{fong2023martingale}.

Proposition \ref{prop:P_N_convergence} quantifies how much the random limit $P_{\infty, x}$ fluctuates around $P_0$, our initial point estimate of the PPD. The spread of these fluctuations characterizes the spread of the martingale posterior for $\theta(P(\cdot | x))$. Consequently, the contraction rate of the posterior is $n^{-\beta + 1/2}$.

\subsection{The role of $\beta$} \label{sec:rolebeta}

The case $\beta = 1$ in \eqref{eq:lr_para} corresponds to the choice $\alpha_i \sim (i + 1)^{-1}$ proposed by \citet{fong2023martingale} in the unconditional context with contraction rate of  $n^{-1/2}$.
Contraction rates for nonparametric regression problems are typically much slower than $n^{-1/2}$ \citep[][Chapter 9]{ghosal2017fundamentals}, however. To accommodate such slow contraction rates, it is therefore crucial to pick a learning rate with $\beta < 1$. The choice $\beta = 1/2 + 2 / (d + 4)$ corresponds to the best possible contraction rate we can expect in a nonparametric regression problem with $d$ features and twice continuously differentiable regression curve.
In nonparametric regression problems, there is, however, an unavoidable estimation bias that is also reflected in the posterior. To account for this bias and achieve asymptotically correct coverage, Bayesian credible intervals have to be inflated, typically using a slowly diverging blow-up factor  \citep{szabo2015frequentist}. 
This motivates a slightly smaller choice for $\beta$, such as
    $\beta = 1/2 + 2 / (1.1d + 4),$
leading to slightly wider credible intervals.

\subsection{The role of initialization} \label{sec:init}

The following corollary clarifies the role of the initialization in our algorithm.

\begin{corollary} \label{cor:init}
Let $P_{0, 1}$ and $P_{0, 2}$ be two PPDs to initialize our algorithm with, and $P_{\infty, x, 1}$ and $P_{\infty, x, 2}$ be the corresponding limit posteriors. For any $x, y$, it holds with probability at least $1 - 2\delta$:
    $$|P_{\infty, x, 1}(y) - P_{\infty, x, 2}(y)|
    \geq |P_{0, 1}(y) - P_{0, 2}(y)| - O(n^{-\beta + 1/2}).$$
\end{corollary}
This is a direct result of \cref{prop:P_N_convergence} and shows that, up to the contraction rate of the posteriors, the difference in martingale posteriors is as large as the difference in the initializations. Hence, 
the choice of initialization is the dominating factor
for the difference in the martingale posteriors.
Our approach, therefore, relies crucially on the quality and calibration of the initial PPD estimate, making full use of the excellent empirical performance of modern PFNs. It then adds an additional computational layer that allows for disentangling the epistemic uncertainty about a predictive summary (e.g., a conditional quantile) from the aleatoric uncertainty in the label $Y$.




\subsection{Computation} \label{sec:computation}

\begin{algorithm}[t] 
\caption{Approximate Martingale Posteriors (AMP)}
\begin{algorithmic}[1] \label{algo:MP}
\STATE \textbf{Input:} Estimated $\wh \PPD(y|x)$ obtained from the PFN.
\FOR{$b = 1, \dots, B$}
    \STATE Initialize $P_0^{(b)} \gets \wh \PPD(y|x)$.
    \FOR{$k = 1, \dots, N$}
        \STATE Sample $y_{n+k}^{(b)} \sim P_{k - 1}^{(b)}$.
        \STATE Update $(P_{k - 1}^{(b)}, y_{n+k}^{(b)}) \to P_{k}^{(b)}$ as in (\ref{eq:gc_update}) with $\alpha_i$ as in \eqref{eq:lr_fixed}.
    \ENDFOR
    \STATE Compute $\wh P_{N}^{(b)}(y) = \frac{1}{N} \sum_{i=1}^N \mathbf{1}\Bigl\{y_{n +i}^{(b)} \le y\Bigr\}$.
    \STATE Set $\theta^{(b)}(x) \gets \theta\bigl( \wh P_{N}^{(b)}\bigr)$.
\ENDFOR
\STATE Compute the estimated Martingale Posterior:\\
$\wh \Pi\Bigl(\theta(x) \in A | \Dcal_n \Bigr) = \frac{1}{B} \sum_{b=1}^B \mathbf{1}\Bigl\{\theta^{(b)}(x) \in A\Bigr\}$.
\end{algorithmic}
\end{algorithm}

The theoretical analysis above assumes that we run the update schemes indefinitely.
In practice, we can only sample finite sequences, which incurs a truncation error. A closer inspection of the proof of \cref{prop:P_N_convergence} reveals that the posterior spread is determined by the square root of the quantity $S_{n, N}^2 := \sum_{i = n}^{n + N-1} \alpha_{i}^2$.
For $N < \infty$, the posterior spread is too small by the multiplicative factor
\begin{align*}
    C_{n, N} = S_{n, N} / S_{n, \infty} \approx \left[1 - \left(1 + N/ n\right)^{1-2\beta}\right]^{1/2}.
\end{align*}
Even for a relatively large number of forward samples $N$, this can be detrimental. For example, with $n = 200$, $N=1000$, and $d = 10$, we get $C_{n, N} \approx 0.62$ and the credible intervals are almost 40\% too small. Knowing this, we can implement a simple fix by appropriately blowing up the intervals, achieved by multiplying the learning rate by $C_{n, N}^{-1}$. Our proposed learning rate is therefore
\begin{align} \label{eq:lr_fixed}
    \alpha_{i} = \min\{1, C_{n, N}^{-1}  2^\beta (i + 1)^{-\beta}\}, \quad \beta = 1/2 + 2 / (1.1d + 4).
\end{align}
The procedure is summarized in \cref{algo:MP}.

The run-time complexity of the algorithm is $O(BN)$, where $B$ is the number of replications, and $N$ is the length of one `chain'. 
The computations are independent across the outer loop ($b = 1, \dots, B$) and straightforward to parallelize or vectorize; the inner loop ($k = 1, \dots, N$) must be run sequentially. Except for the initial PPD computation, the runtime is independent of the training set size $n$.

\section{Numerical experiments}
\label{sec:experiments}

To demonstrate the efficacy of our approach, we conduct various experiments. 
Code is made available at \url{https://anonymous.4open.science/r/UQ_for_PFNs}.
Our experiments use both TabPFN and TabICL to demonstrate the algorithm's versatility.
Unless stated otherwise, AMP uses $N=50$ forward samples and $B=50$ independent `chains'.

\subsection{Simulation study} \label{sec:sim}

We first assess the calibration of the algorithm in a simulation study using Bayesian additive model DGPs as ground-truth. We generate the true underlying feature effects $f_j(x_j) = B(x_j)^\top\theta_j$ for feature $x_j$ by drawing parameters $\theta_j\sim\mathcal{N}(0, 2 I)$ and using those parameters as spline coefficients for a B-spline basis $B_j(x_j)\in\mathbb{R}^{20}$ of degree 3 with 20 basis functions. A training data set of size $n \in \{50, 100, 200, 400, 800\}$ is then generated by the law $Y|x,\theta\sim\mathcal{N}(\sum_{j=1}^J f_j(x_j), 1)$. We use $d \in \{1, 10, 20\}$ features from which $J\in\{\lceil d/2 \rceil, d\}$ have a non-zero signal while the rest act as noise features with no contribution to $Y|x$. Using the true DGP as a Bayesian prior, the resulting posterior is analytically available and can be used to construct credible intervals for conditional quantiles (referred to as \emph{Optimal}). This allows for comparing the quality of epistemic uncertainty to a gold-standard baseline.  For each $(n, d, J)$ combination, we simulate 20 data sets and evaluate all methods on 100 random test points.

\begin{figure}[t]
    \centering
    \includegraphics[width=\columnwidth]{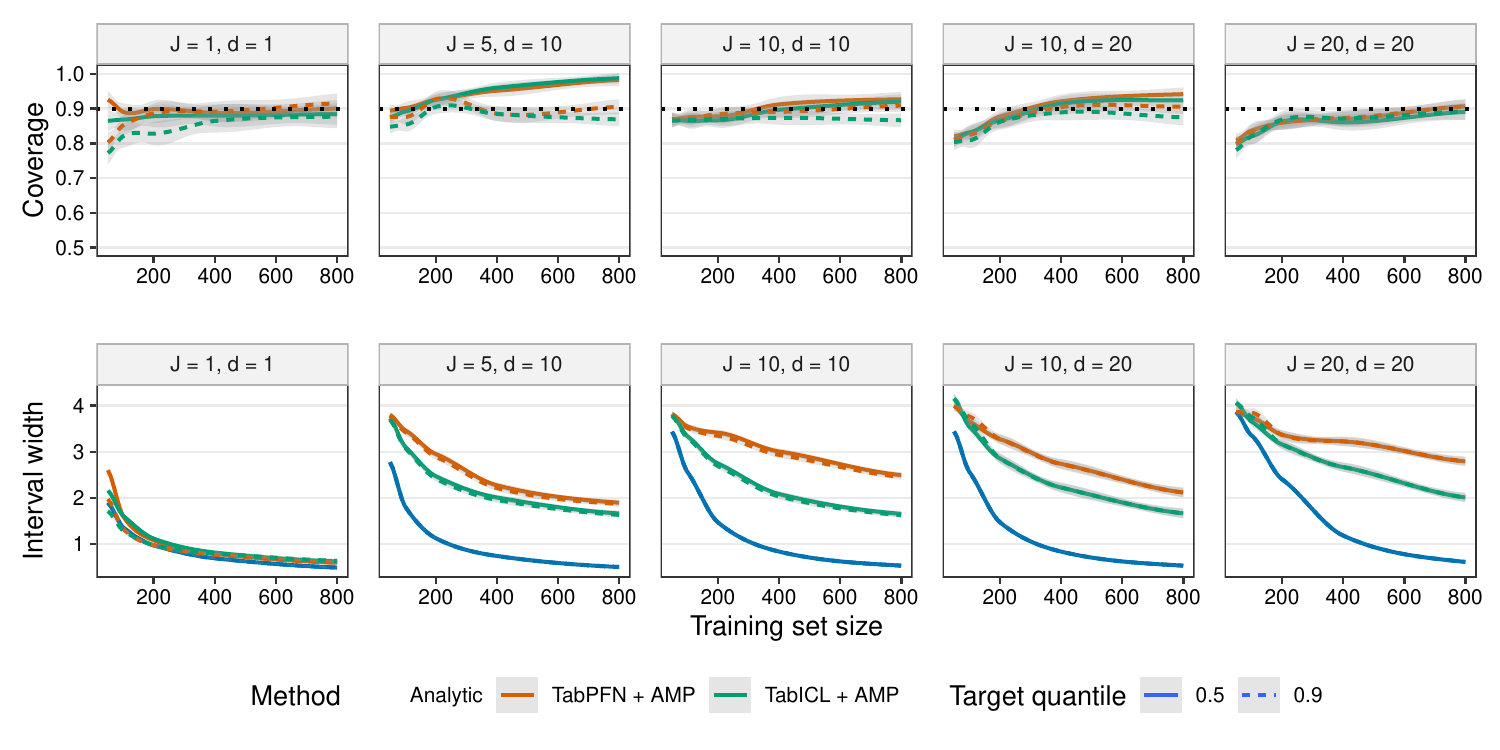}
    \caption{Coverage (target: 90\%) and CI width for data simulated from a Bayesian additive model, comparing our approximate martingale posterior (AMP) credible sets for TabPFN and TabICL with analytic credible sets using the true DGP as prior. All widths are scaled by $d^{-1/2}$ for better visibility.}
    \label{fig:sim_study}
\end{figure}

\paragraph{Results} \cref{fig:sim_study} shows the results of our simulation experiment with average coverage in the top row and average interval width in the bottom row. First, we observe that all methods have approximately correct coverage, with the lowest points at around 80\% for $n=50$ in some settings. For $J < d$, the PFNs appear over-conservative for the median estimates. Our routine is unaware that some features are irrelevant, but the PFN may exploit this to yield better-than-expected point estimates. The interval widths decrease with the training set size, as expected. The widths for the analytic method decrease much faster for $d > 1$, because this method exploits full knowledge of additivity, irrelevant features, and the noise variance. The AMP method also decreases width with training set size, but appropriately adapts to the actual uncertainty in the PFN estimators.

\subsection{Ablation studies} \label{sec:exp_ablations}

\begin{figure}[t]
    \centering
    \includegraphics[width=\linewidth]{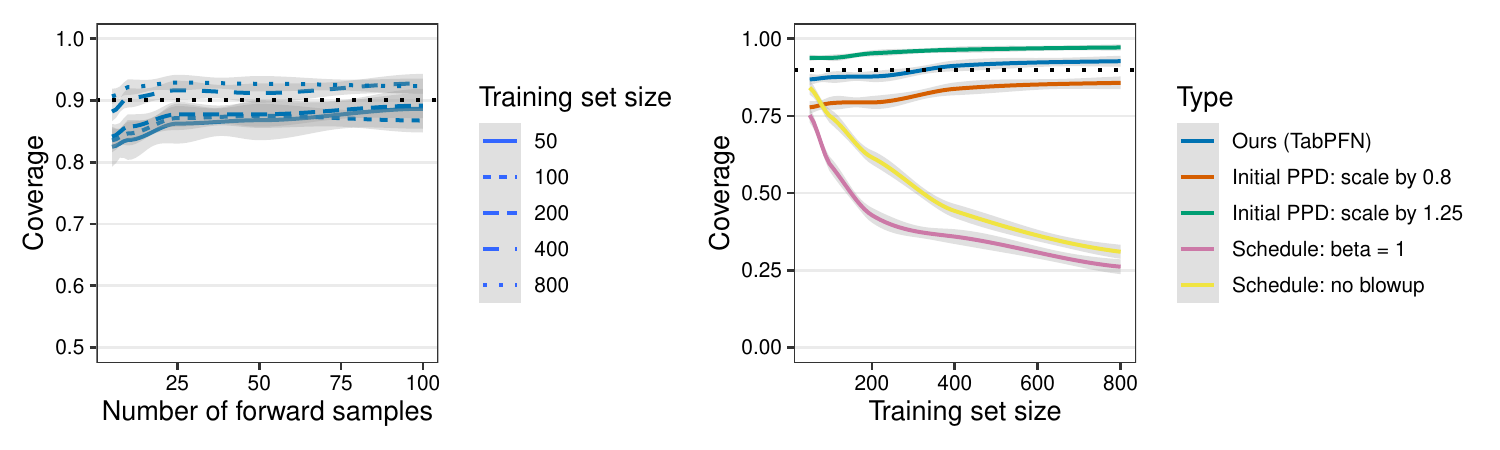}
    \caption{Ablation studies using TabPFN on simulated data with $d=10, J=10$. The left panel shows coverage for a given number of forward samples $N$; the right panel varies the initial PPD estimate and the learning rate schedule.}
    \label{fig:ablation_main}
\end{figure}

In \cref{fig:ablation_main}, we assess the role of other factors affecting performance using TabPFN on simulated data with $d=10$ and $J=10$ as an example; extended results are given in Appendix \cref{app:ablation_add}.

\paragraph{Number of forward samples} Our main experiments used $N=50$ forward samples as default. The left panel of \cref{fig:ablation_main} shows the effect of this parameter on the coverage.
The coverage appears to stabilize already at around $N=20$ forward samples thanks to the blow-up factor introduced in \cref{sec:computation}, so $N=50$ is already a conservative choice. For $d = 20$, $N=50$ appears appropriate (see \cref{app:ablation_add}).

\paragraph{Initialization} 
To corroborate the need for a well-calibrated initial PPD estimate, we compress and widen the PPD (relative to the conditional median) by factors of 0.8 and 1.25, respectively. In the right panel of \cref{fig:ablation_main}, we see a direct effect on coverage: the compressed initial PPD yields lower coverage, while the widened version yields higher coverage.

\paragraph{Learning rate schedule}
The right panel of \cref{fig:ablation_main} further shows two alternatives to our learning rate schedule \eqref{eq:lr_fixed}. As expected from our theoretical arguments in \cref{sec:rolebeta} and \cref{sec:computation}, using $\beta = 1$ rather than $\beta = 1/2 + 2 /(1.1d + 4)$ or removing the blow-up factor $C_{n, N}^{-1}$ from the schedule leads to severe undercoverage. 

\subsection{Benchmarks on real data} \label{sec:benchmark}

\begin{table} 
\caption{Benchmark results for 90\%-quantile regression on various UCI data sets, reporting averages over ten random 20-80 train-test splits with $\pm 2\mathrm{se}$.  The target coverage is 90\%.}
\centering
\footnotesize

{\setlength\tabcolsep{4.5pt}\begin{tabular}{cclrrrrrrrr}
\toprule
 &  & Metric & airfoil & boston & concrete & diabetes & energy & fish & forest & real\\
\midrule
 &  & Coverage & 0.79{\tiny $\pm$0.01} & 0.87{\tiny $\pm$0.04} & 0.86{\tiny $\pm$0.02} & 0.94{\tiny $\pm$0.04} & 0.89{\tiny $\pm$0.01} & 0.89{\tiny $\pm$0.03} & 0.99{\tiny $\pm$0.01} & 0.88{\tiny $\pm$0.03}\\
 &  & Width & 0.56{\tiny $\pm$0.03} & 1.01{\tiny $\pm$0.06} & 0.88{\tiny $\pm$0.02} & 2.31{\tiny $\pm$0.10} & 0.15{\tiny $\pm$0.01} & 1.61{\tiny $\pm$0.09} & 2.37{\tiny $\pm$0.09} & 2.15{\tiny $\pm$0.35}\\
 & \multirow{-3}{*}{\centering\arraybackslash \raisebox{-0.5\height}{\rotatebox[origin=c]{90}{TabPFN}}} & Time (s) & 20.5{\tiny $\pm$0.2} & 7.1{\tiny $\pm$0.8} & 13.7{\tiny $\pm$1.1} & 5.7{\tiny $\pm$0.4} & 9.7{\tiny $\pm$0.6} & 12.6{\tiny $\pm$0.8} & 7.6{\tiny $\pm$0.6} & 5.1{\tiny $\pm$0.1}\\
\cmidrule(lr){2-11}
 &  & Coverage & 0.72{\tiny $\pm$0.01} & 0.89{\tiny $\pm$0.03} & 0.85{\tiny $\pm$0.02} & 0.89{\tiny $\pm$0.05} & 0.84{\tiny $\pm$0.02} & 0.91{\tiny $\pm$0.03} & 0.86{\tiny $\pm$0.05} & 0.87{\tiny $\pm$0.03}\\
 &  & Width & 0.55{\tiny $\pm$0.02} & 1.03{\tiny $\pm$0.06} & 0.82{\tiny $\pm$0.03} & 2.26{\tiny $\pm$0.08} & 0.13{\tiny $\pm$0.01} & 1.66{\tiny $\pm$0.09} & 2.90{\tiny $\pm$0.11} & 2.43{\tiny $\pm$0.29}\\
\multirow{-6}{*}[0.5\dimexpr\aboverulesep+\belowrulesep+\cmidrulewidth]{\centering\arraybackslash \raisebox{-0.5\height}{\rotatebox[origin=c]{90}{AMP}}} & \multirow{-3}{*}{\centering\arraybackslash \raisebox{-0.5\height}{\rotatebox[origin=c]{90}{TabICL}}} & Time (s) & 20.6{\tiny $\pm$0.2} & 6.3{\tiny $\pm$0.4} & 13.8{\tiny $\pm$1.0} & 5.8{\tiny $\pm$0.3} & 10.1{\tiny $\pm$0.9} & 13.3{\tiny $\pm$1.1} & 7.1{\tiny $\pm$0.5} & 5.0{\tiny $\pm$0.0}\\
\midrule[1.2pt]
 &  & Coverage & 0.60{\tiny $\pm$0.02} & 0.71{\tiny $\pm$0.03} & 0.57{\tiny $\pm$0.01} & 0.85{\tiny $\pm$0.04} & 0.77{\tiny $\pm$0.02} & 0.70{\tiny $\pm$0.03} & 0.90{\tiny $\pm$0.05} & 0.70{\tiny $\pm$0.03}\\
 &  & Width & 0.57{\tiny $\pm$0.03} & 0.77{\tiny $\pm$0.07} & 0.63{\tiny $\pm$0.02} & 1.12{\tiny $\pm$0.07} & 0.17{\tiny $\pm$0.01} & 0.98{\tiny $\pm$0.07} & 1.75{\tiny $\pm$0.13} & 0.59{\tiny $\pm$0.06}\\
 & \multirow{-3}{*}{\centering\arraybackslash \raisebox{-0.5\height}{\rotatebox[origin=c]{90}{TabPFN}}} & Time (s) & 76.3{\tiny $\pm$30.1} & 17.4{\tiny $\pm$5.9} & 53.0{\tiny $\pm$35.3} & 41.9{\tiny $\pm$23.9} & 67.0{\tiny $\pm$25.1} & 53.4{\tiny $\pm$14.6} & 33.7{\tiny $\pm$8.5} & 10.3{\tiny $\pm$1.1}\\
\cmidrule(lr){2-11}
 &  & Coverage & 0.50{\tiny $\pm$0.02} & 0.45{\tiny $\pm$0.03} & 0.43{\tiny $\pm$0.02} & 0.81{\tiny $\pm$0.04} & 0.80{\tiny $\pm$0.02} & 0.40{\tiny $\pm$0.04} & 0.72{\tiny $\pm$0.09} & 0.55{\tiny $\pm$0.04}\\
 &  & Width & 0.57{\tiny $\pm$0.03} & 0.69{\tiny $\pm$0.07} & 0.55{\tiny $\pm$0.01} & 0.84{\tiny $\pm$0.04} & 0.10{\tiny $\pm$0.01} & 0.69{\tiny $\pm$0.05} & 1.21{\tiny $\pm$0.12} & 0.52{\tiny $\pm$0.05}\\
\multirow{-6}{*}[0.5\dimexpr\aboverulesep+\belowrulesep+\cmidrulewidth]{\centering\arraybackslash \raisebox{-0.5\height}{\rotatebox[origin=c]{90}{Bootstrap}}} & \multirow{-3}{*}{\centering\arraybackslash \raisebox{-0.5\height}{\rotatebox[origin=c]{90}{TabICL}}} & Time (s) & 70.0{\tiny $\pm$30.1} & 65.7{\tiny $\pm$29.8} & 54.7{\tiny $\pm$26.9} & 41.7{\tiny $\pm$13.1} & 39.0{\tiny $\pm$17.2} & 38.2{\tiny $\pm$19.9} & 22.5{\tiny $\pm$5.2} & 13.5{\tiny $\pm$0.2}\\
\bottomrule
\end{tabular}}

\label{tab:benchmark_q90}
\end{table}

We next assess our method on a suite of real-world data from the UCI repository \citep{Dua.2019}; see Appendix \ref{app:data_desc} for a description of the data sets.
Since the true conditional quantile is not known on real data sets, we use the following surrogate procedure.
We use 20\% of the data for training and compute predicted quantiles along with credible intervals for the held-out test data. We then check whether the predicted quantile from a separate PFN trained and evaluated only on the test data falls within the intervals. Since the latter PFN is trained on a much larger data set, it serves as a reasonable `oracle` surrogate for the true conditional quantile.
While the training set sizes are rather small, this provides an ideal benchmark for PFNs, which remain limited in their scalability to large datasets. 



We compare against a bootstrap approach as a frequentist natural baseline, as it also assesses epistemic uncertainty for PFN estimates without knowledge of the pre-training pipeline. 
The bootstrap draws samples with replacement from the training dataset, uses TabPFN to make quantile predictions on the test set, and computes an empirical confidence interval across $B=50$ replications.  In all cases, we construct a 90\% credible/confidence interval (CI). 

\paragraph{Results} \cref{tab:benchmark_q90} shows the results of our coverage benchmark for predicting the conditional 90\%-quantile; results for median regression are given in Appendix \ref{app:benchmark_add}. Our AMP method provides approximately correct coverage across most datasets, though it can be slightly overconservative for TabPFN and conditional medians. The bootstrap produces narrower intervals and undercovers in most cases for the same reason that requires inflating credible sets: it ignores the unavoidable estimation bias in nonparametric regression problems \citep{ hall2013simple}.
Notably, the bootstrap is several times slower than our AMP method.
While AMP fits the PFN only once on the training data, the bootstrap requires $B$ refits, each costing $O(n^2)$. 
The AMP method computes CIs for all test points in a few seconds, making it an ideal companion for modern PFNs.

\section{Discussion} \label{sec:discussion}

This work proposes an efficient and principled Bayesian uncertainty quantification method for estimates derived from prior-data fitted networks. The method provides posterior uncertainty for predictive summaries, such as conditional means or quantiles, which is not available from the PFN posterior predictive distribution alone. Compared with more classical approaches such as deep ensembles \citep{lakshminarayanan_2017_SimpleScalablea}, the proposed procedure is particularly well suited to pretrained models: it requires only one PFN evaluation on the training data and does not need access or modifications of the pre-training pipeline. By providing tuning-free, well-calibrated credible intervals in seconds, it serves as an ideal uncertainty layer playing to the strengths of modern PFNs.

\paragraph{Limitations}
Our theory and experiments currently focus on iid tabular prediction problems. Although the same martingale posterior idea may be applicable beyond this setting, extensions to non-tabular or dependent data require additional assumptions and analysis. 
A second limitation is that the current implementation computes posterior credible intervals for each query point separately. Thus, the computational cost scales with the size of the test set, although the method is sufficiently fast in our experiments to evaluate hundreds of query points within seconds. 

The pointwise nature of the current procedure also means that it does not provide uniform credible sets for joint posterior inference.
A full joint posterior would require modeling the dependence across feature values, so the distribution of $x_{1:n}$ can no longer be ignored. \citet{fong2023martingale} proposed a general joint update of the PPDs over all values of $x$, but this update is intractable in practice, and the proposed heuristic simplifications are neither especially simple nor theoretically justified. A possible direction is to combine the marginal posteriors $P_{\infty,x_1}, \dots, P_{\infty,x_K}$ through a dependence model, for example, a multivariate Gaussian copula with a covariance kernel depending on distances between query points, or the more flexible vine-copula construction proposed by \citet{huk2024quasi}. Developing such joint posterior extensions is a promising direction for future work.
%

\bibliography{references}
\bibliographystyle{apalike}




\appendix

\section{Proof of Proposition \texorpdfstring{\ref{prop:P_N_convergence}}{\ref*{prop:P_N_convergence}}} \label{sec:proof}

Let $P_0$ be an arbitrary probability distribution on $\R$.
We first show that $P_M(y), M \ge 0,$ is a martingale.
It holds
    \begin{align*}
        &\quad \E[P_{M}(y) | y_{(n + 1):(n + M - 1)}] \\
        &= (1 - \alpha_{n + M -1}) P_{M-1}(y) + \alpha_{n + M -1} \E_{y_{n + M} \sim P_{M - 1}}[H_\rho(P_{M - 1}(y), P_{M - 1}(y_{n + M})].
    \end{align*}
    By the probability integral transform, it holds $P_{M - 1}(y_{n + M}) \sim \mathrm{Uniform}[0, 1]$. Thus, 
    \begin{align*}
        \E_{y_{n + M} \sim P_{M - 1}}[H_\rho(P_{M - 1}(y), P_{M - 1}(y_{n + M})] = \int H_\rho(P_{M - 1}(y), u) du = P_{M - 1}(y),
    \end{align*}
    by the properties of the Gaussian copula. Hence, 
    \begin{align*}
        \E[P_{M}(y) | y_{(n + 1):(n + M - 1)}] = P_{M - 1}(y),
    \end{align*}
    which implies that $P_M(y)$ is a martingale adapted to the filtration $\sigma(y_{(n+1):(n+M)})_{M \ge 1}$. 
    
    Next, we apply the uniform-in-time version of the Hoeffding-Azuma inequality given in Corollary 1 (a) of \citet{howard2020time}.
    It holds
    \begin{align*}
        |P_M(y) - P_{M - 1}(y)| \le \alpha_{n + M -1} \le 2^\beta (n + M + 1)^{-\beta}, \quad \text{for all } y \in \R,
    \end{align*}
    and, for all $M \ge 0$,
    \begin{align*}
       \sum_{i = 0}^M \alpha_{n + i}^2  \le \sum_{i = 0}^\infty \alpha_{n + i}^2 \le 2^{2\beta} \int_{0}^\infty (n + t)^{-2\beta} dt \le C_\beta   n^{1-2\beta},
    \end{align*}
    where $C_\beta = 2^{2\beta}/(2\beta - 1)$.
    
    Now Corollary 1 (a) of \citet{howard2020time} with $m = C_\beta n^{1-2\beta}$ yields 
    \begin{align*}
        \Pr(\exists M \ge 0\colon |P_M(y) - P_{0}(y)| \ge \epsilon) \le 2 \exp\left(-\frac{\epsilon^2}{2 m } \right) = 2\exp\left(-\frac{\epsilon^2 n^{2\beta - 1} }{2C_\beta} \right).
    \end{align*}
    Setting $\eps =  \sqrt{2C_\beta \log(2/\delta)} n^{1/2 - \beta}$ gives,
    \begin{align*}
        \Pr(\exists M \ge 0\colon |P_M(y) - P_{0}(y)| \ge \epsilon) \le \delta.
    \end{align*}
    Now the claim follows since $P_{M}(y) \to P_{\infty, x}(y)$ almost surely (\cref{prop:existence}).
    \qedsymbol

\section{Further experimental results and details} \label{app:experiments}

\FloatBarrier
\subsection{Additional ablation results}
\label{app:ablation_add}

\begin{figure}[h]
    \centering
    \includegraphics[width=\columnwidth]{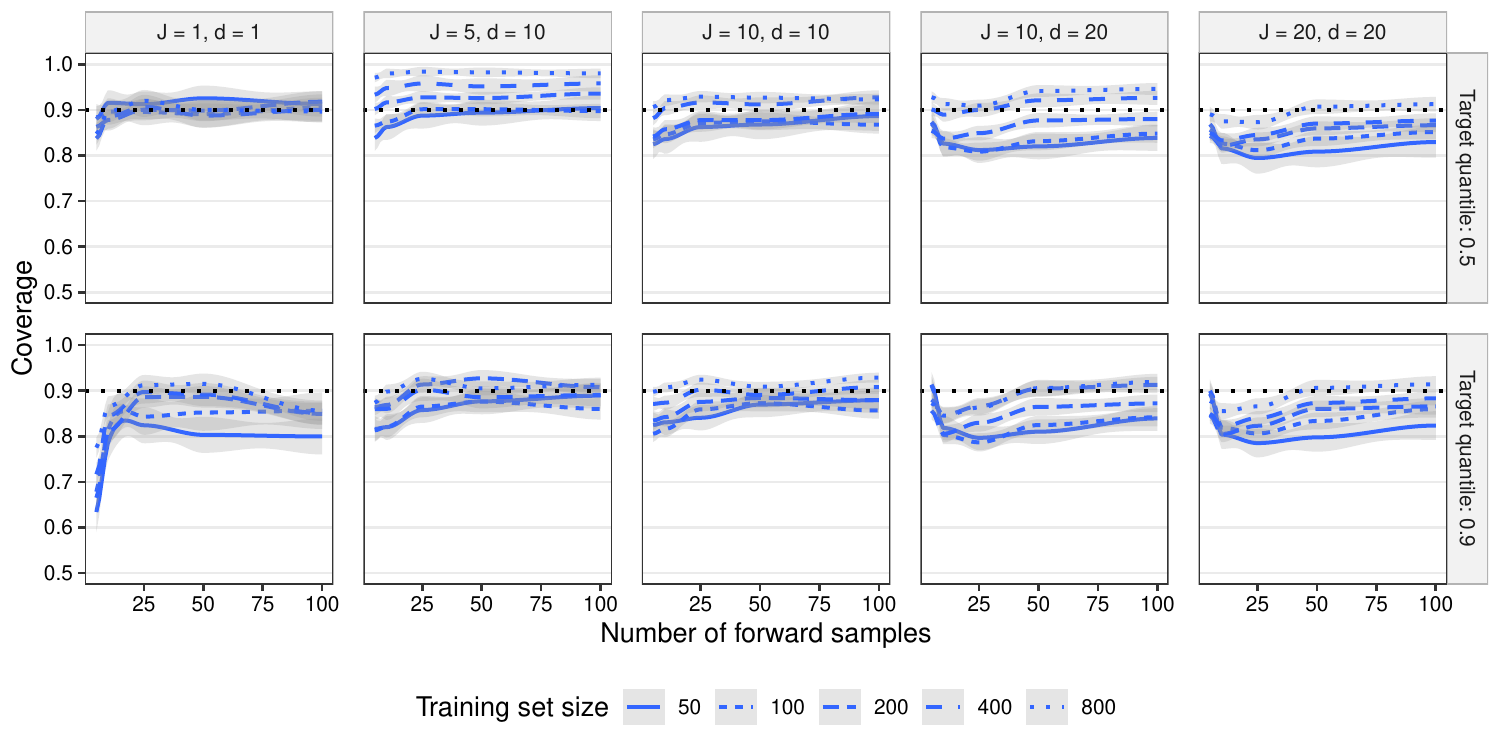}
    \caption{Ablation results for the number of forward samples $N$ on all simulation settings.}
    \label{fig:ablation_T}
\end{figure}

\begin{figure}[h]
    \centering
    \includegraphics[width=\columnwidth]{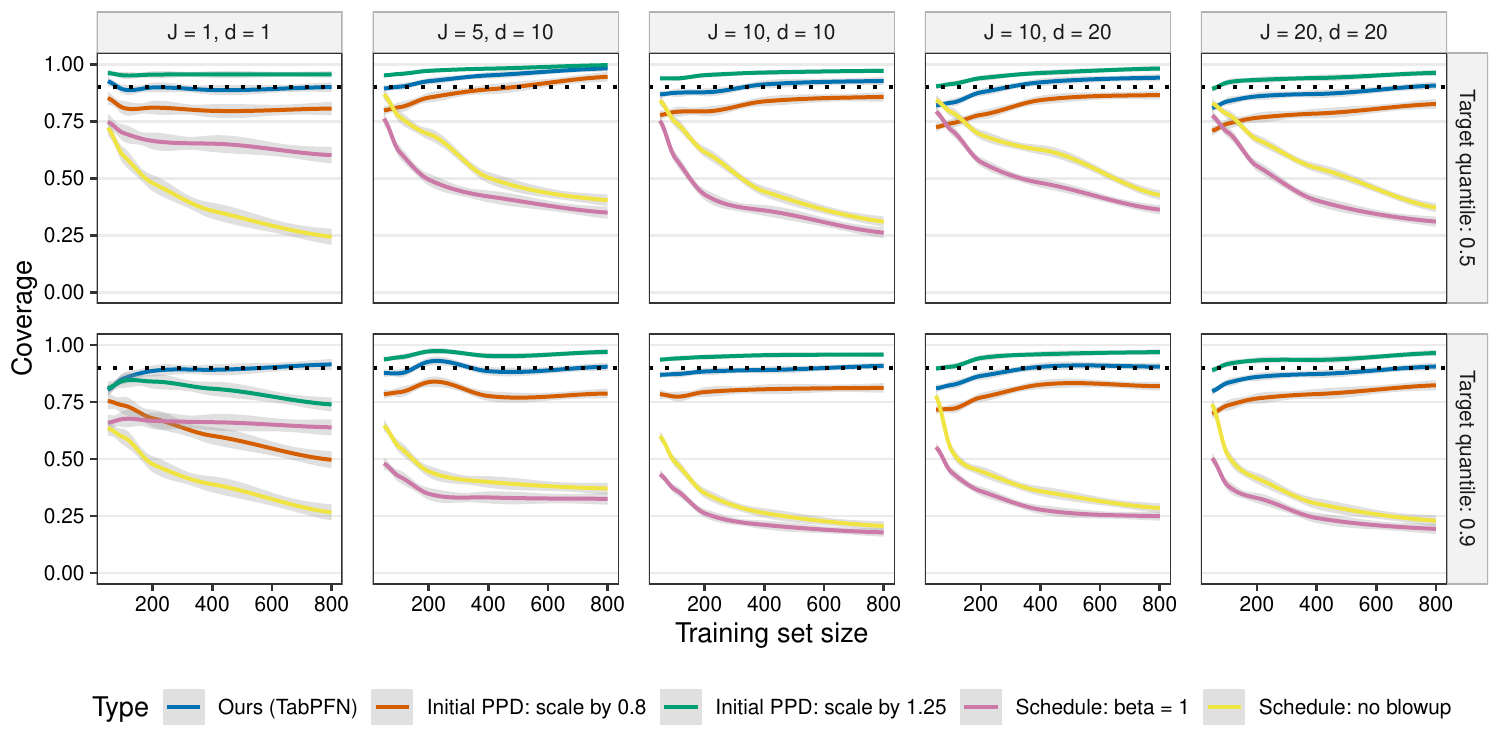}
    \caption{Ablation results for initialization and schedule on all simulation settings.}
    \label{fig:ablation_schedule}
\end{figure}

\FloatBarrier

\subsection{Real data sets} \label{app:data_desc}

We use the following data sets from the UCI repository \citep{Dua.2019}:

\begin{itemize}
    \item \texttt{airfoil} ($n=1503$, $d=5$), 
    \item \texttt{boston}  ($n=252$, $d=15$),
    \item \texttt{concrete} ($n=1030$, $d=8$, \citealp{Yeh.1998}),  
    \item \texttt{diabetes} ($n=442$, $d=10$, \citealp{Efron.2004}), 
    \item \texttt{energy} ($n=768$, $d=8$, \citealp{Tsanas.2012}),
    \item \texttt{fish} ($n=908$, $d=6$), 
    \item \texttt{forest\_fire} ($n=516$, $d=13$),
    \item \texttt{real} ($n=413$, $d=7$).
\end{itemize}

\subsection{Additional benchmark results}
\label{app:benchmark_add}

\begin{table}[h!]
\caption{Benchmark results for median regression on various UCI data sets, reporting averages over ten random 20-80 train-test splits with $\pm 2\mathrm{se}$.  The target coverage is 90\%.}
\centering
\footnotesize

{\setlength\tabcolsep{4.5pt}\begin{tabular}{cclrrrrrrrr}
\toprule
 &  & Metric & airfoil & boston & concrete & diabetes & energy & fish & forest & real\\
\midrule
 &  & Coverage & 0.86{\tiny $\pm$0.01} & 0.93{\tiny $\pm$0.03} & 0.90{\tiny $\pm$0.01} & 1.00{\tiny $\pm$0.00} & 0.95{\tiny $\pm$0.01} & 0.98{\tiny $\pm$0.01} & 0.70{\tiny $\pm$0.24} & 0.98{\tiny $\pm$0.01}\\
 &  & Width & 0.65{\tiny $\pm$0.03} & 1.01{\tiny $\pm$0.06} & 0.95{\tiny $\pm$0.03} & 2.21{\tiny $\pm$0.10} & 0.16{\tiny $\pm$0.01} & 1.72{\tiny $\pm$0.08} & 2.37{\tiny $\pm$0.09} & 3.08{\tiny $\pm$0.28}\\
 & \multirow{-3}{*}{\centering\arraybackslash \raisebox{-0.5\height}{\rotatebox[origin=c]{90}{TabPFN}}} & Time (s) & 20.4{\tiny $\pm$0.1} & 6.1{\tiny $\pm$0.3} & 13.7{\tiny $\pm$0.7} & 5.9{\tiny $\pm$0.5} & 9.1{\tiny $\pm$0.1} & 11.3{\tiny $\pm$0.3} & 6.8{\tiny $\pm$0.6} & 5.3{\tiny $\pm$0.3}\\
\cmidrule(lr){2-11}
 &  & Coverage & 0.84{\tiny $\pm$0.02} & 0.88{\tiny $\pm$0.03} & 0.89{\tiny $\pm$0.02} & 1.00{\tiny $\pm$0.00} & 0.93{\tiny $\pm$0.01} & 0.94{\tiny $\pm$0.01} & 0.96{\tiny $\pm$0.03} & 0.96{\tiny $\pm$0.02}\\
 &  & Width & 0.64{\tiny $\pm$0.02} & 1.03{\tiny $\pm$0.06} & 0.89{\tiny $\pm$0.03} & 2.15{\tiny $\pm$0.08} & 0.14{\tiny $\pm$0.01} & 1.75{\tiny $\pm$0.07} & 2.90{\tiny $\pm$0.11} & 3.08{\tiny $\pm$0.28}\\
\multirow{-6}{*}[0.5\dimexpr\aboverulesep+\belowrulesep+\cmidrulewidth]{\centering\arraybackslash \raisebox{-0.5\height}{\rotatebox[origin=c]{90}{AMP}}} & \multirow{-3}{*}{\centering\arraybackslash \raisebox{-0.5\height}{\rotatebox[origin=c]{90}{TabICL}}} & Time (s) & 21.1{\tiny $\pm$0.5} & 6.2{\tiny $\pm$0.3} & 14.2{\tiny $\pm$1.2} & 6.0{\tiny $\pm$0.4} & 9.3{\tiny $\pm$0.2} & 11.2{\tiny $\pm$0.2} & 6.7{\tiny $\pm$0.5} & 5.2{\tiny $\pm$0.4}\\
\midrule[1.2pt]
 &  & Coverage & 0.67{\tiny $\pm$0.01} & 0.71{\tiny $\pm$0.03} & 0.62{\tiny $\pm$0.01} & 0.86{\tiny $\pm$0.04} & 0.78{\tiny $\pm$0.01} & 0.71{\tiny $\pm$0.02} & 0.87{\tiny $\pm$0.07} & 0.75{\tiny $\pm$0.04}\\
 &  & Width & 0.48{\tiny $\pm$0.02} & 0.55{\tiny $\pm$0.03} & 0.51{\tiny $\pm$0.01} & 1.14{\tiny $\pm$0.04} & 0.11{\tiny $\pm$0.01} & 0.79{\tiny $\pm$0.03} & 1.09{\tiny $\pm$0.10} & 1.09{\tiny $\pm$0.12}\\
 & \multirow{-3}{*}{\centering\arraybackslash \raisebox{-0.5\height}{\rotatebox[origin=c]{90}{TabPFN}}} & Time (s) & 83.1{\tiny $\pm$32.5} & 35.0{\tiny $\pm$17.0} & 56.2{\tiny $\pm$40.3} & 67.2{\tiny $\pm$34.1} & 57.9{\tiny $\pm$27.0} & 17.8{\tiny $\pm$4.2} & 20.5{\tiny $\pm$4.5} & 15.9{\tiny $\pm$4.9}\\
\cmidrule(lr){2-11}
 &  & Coverage & 0.69{\tiny $\pm$0.02} & 0.58{\tiny $\pm$0.02} & 0.58{\tiny $\pm$0.01} & 0.82{\tiny $\pm$0.03} & 0.80{\tiny $\pm$0.02} & 0.48{\tiny $\pm$0.02} & 0.52{\tiny $\pm$0.11} & 0.66{\tiny $\pm$0.02}\\
 &  & Width & 0.51{\tiny $\pm$0.02} & 0.50{\tiny $\pm$0.03} & 0.46{\tiny $\pm$0.01} & 0.81{\tiny $\pm$0.02} & 0.09{\tiny $\pm$0.01} & 0.53{\tiny $\pm$0.02} & 0.81{\tiny $\pm$0.08} & 0.99{\tiny $\pm$0.13}\\
\multirow{-6}{*}[0.5\dimexpr\aboverulesep+\belowrulesep+\cmidrulewidth]{\centering\arraybackslash \raisebox{-0.5\height}{\rotatebox[origin=c]{90}{Bootstrap}}} & \multirow{-3}{*}{\centering\arraybackslash \raisebox{-0.5\height}{\rotatebox[origin=c]{90}{TabICL}}} & Time (s) & 64.4{\tiny $\pm$27.6} & 32.4{\tiny $\pm$12.5} & 48.3{\tiny $\pm$26.4} & 24.8{\tiny $\pm$5.5} & 48.8{\tiny $\pm$18.9} & 58.0{\tiny $\pm$14.5} & 40.3{\tiny $\pm$11.6} & 33.2{\tiny $\pm$13.2}\\
\bottomrule
\end{tabular}}

 \label{tab:benchmark_q50}
\end{table}

\FloatBarrier

\subsection{Computational environment}
\label{app:compute}

All computations were performed on a user PC with a GeForce RTX 4070 Ti GPU using Python 3.12.7. The total run time of the experiments does not exceed 24 hours. All experiments are run with versions \texttt{tabpfn==2.0.5} (\texttt{n\_estimators=8}) and \texttt{tabicl==2.0.3} (\texttt{n\_estimators=2}).

\end{document}